\documentclass{iucr}              

\usepackage{amsmath} 

\begin{document}                  



\title{Optimization strategies and artefacts of time-involved small angle neutron scattering experiments}


\cauthor[a]{Denis}{Mettus}{denis.mettus@tum.de}

\author[a]{Alfonso}{Chacon}
\author[a,b]{Andreas}{Bauer}
\author[c]{Sebastian}{M\"uhlbauer}
\author[a,b,d]{Christian}{Pfleiderer}

\aff[a]{Physik Department, Technische Universit\"at M\"unchen, Garching, \country{Germany}}
\aff[b]{Centre for QuantumEngineering (ZQE), Technical University of Munich, D-85748 Garching, \country{Germany}}
\aff[c]{Heinz Maier-Leibnitz Zentrum (MLZ), Technische Universit\"at M\"unchen, Garching, \country{Germany}}
\aff[d]{Munich Center for Quantum Science and Technology (MCQST), Technical University of Munich, D-85748 Garching, \country{Germany}}



\maketitle                        

\begin{abstract}
Kinetic small-angle neutron scattering provides access to the microscopic properties of mesoscale systems under slow, periodic perturbations. By interlocking the phases of neutron pulse, sample modulation, and detector signal, so-called Time-Involved Small Angle Neutron scattering Experiments (TISANE) allow to exploit the neutron velocity spread and record data without major sacrifice in intensity at time-scales down to micro-seconds. We review the optimization strategies of TISANE that arise from specific aspects of the process of data acquisition and data analysis starting from the basic principles of operation. Typical artefacts of data recorded in TISANE due to the choice of time-binning and neutron chopper pulse width are illustrated by virtue of the response of the skyrmion lattice in MnSi under periodic changes of the direction of the magnetic field stabilizing the skyrmion lattice. 
\end{abstract}


\section{Introduction}
\\
Small-Angle Neutron Scattering (SANS) is widely used as a probe of mesoscopic length scales up to several hundred nanometers in disciplines as diverse as material science, physics, chemistry, and biology ~\cite{mue_RMP2019, Muehlbauer_NIMPRA297_2016, Dewhurst_JAC49_2016, Wood_JAC51_2018, Kohlbrecher_JAC33_2000, ISIS_SANS2D, Barker_2022_JAC_55, Glinka_1998_JAC_31}. Providing reciprocal space information, SANS is complementary to real-space and surface-sensitive imaging techniques, e.g., scanning-tunneling,  force, transmission-electron, and optical microscopy, or even electron holography. The growing need for information on the dynamic properties of mesoscale structures has motivated time-resolved SANS, resulting, more recently, in the development of so-called Time-Involved Small Angle Neutron scattering Experiments (TISANE)~\cite{Wiedenmann_PRL97_2006, Kippling_PLA372_2008}. Targeting the dynamic response under a periodic drive, TISANE, allow to extend the time resolution of conventional SANS to well below sub-milliseconds at high scattering intensities.

In the light of the rapidly growing scientific and technological interest in mesoscale textures in quantum materials such as superconducting vortex matter, long-wavelength magnetic modulations, or skyrmion and meron lattices, a large number of scientific questions have recently emerged that may be addressed by means of time-resolved SANS studies. For instance, stroboscopic SANS has been used to study the vortex lattice dynamics in superconducting niobium~\cite{mue_PRB2011}, show-casing the determination of the elasticity moduli, as well as vortex lattice relaxation and diffusion. Similarly, TISANE has been used to track the unpinning of the skyrmion lattice in MnSi under periodic changes of field direction~\cite{mue_NJP2016}. The excitation of Ni nanorod colloids under oscillating magnetic fields allowed to identify differences of response as a function of the amplitude and frequency of the driving field~\cite{Bender_Nanoscale_2015}. Recent studies~\cite{Glinka_2020_JAC_53} report utilization of TISANE for the study of hematite spindles in oscillating magnetic fields.

In this paper we review basic aspects of time-resolved SANS, focusing on optimization strategies of TISANE. Starting from the principles of operation, we discuss key elements of the data acquisition and analysis. Considering the example of skyrmion lattice kinetics in MnSi, we illustrate prominent artefacts associated with the choice of time-binning and neutron pulse width. 

\section{Time-resolved small-angle neutron scattering}
\subsection{Time-resolved SANS experiment with a continuous beam}

\begin{figure}
\includegraphics[width=1.0\linewidth]{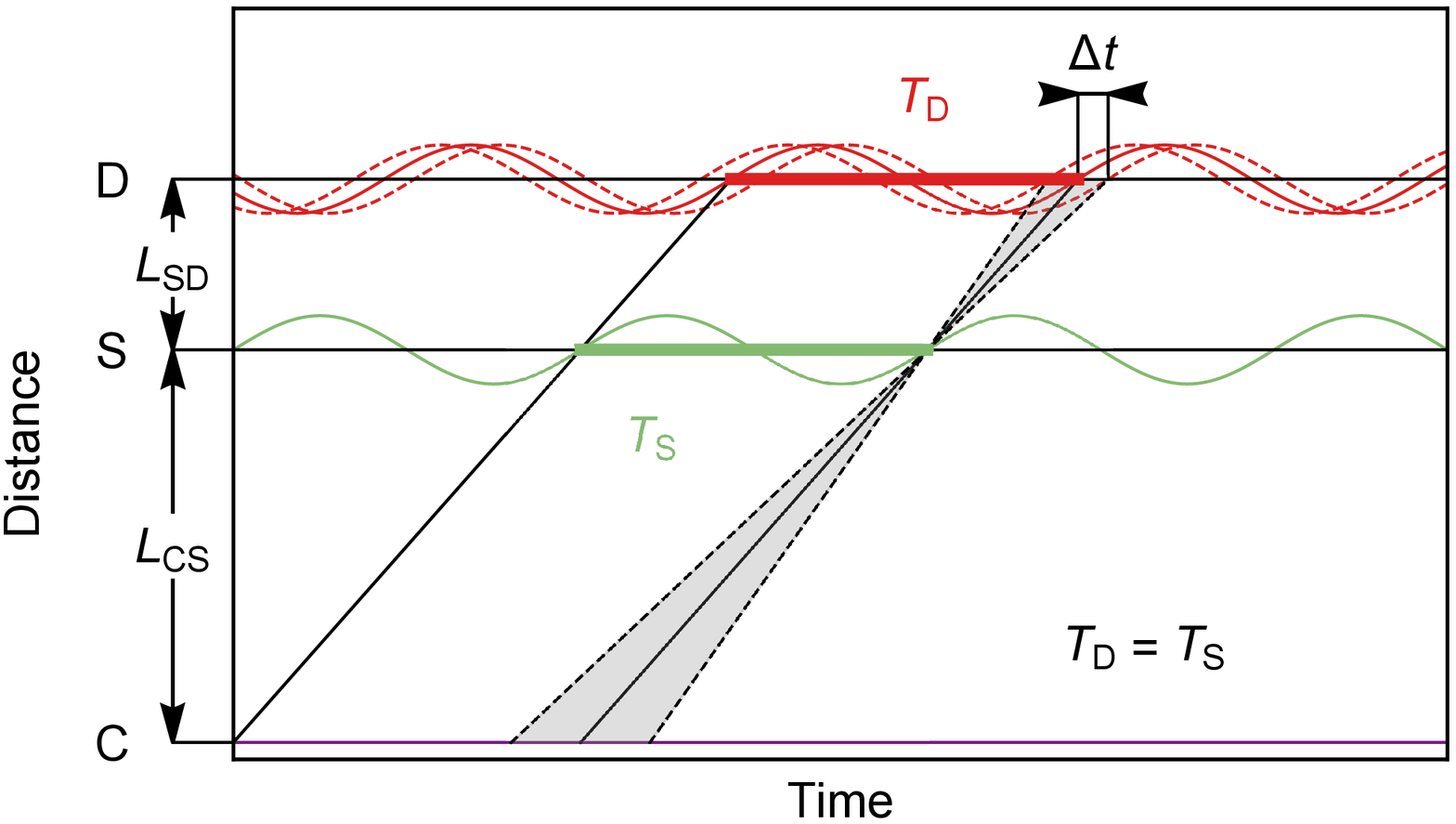}
\caption{Time versus distance diagram of SANS with a continuous neutron beam. The velocity selector (C) generates a continuous beam with wavelength $\lambda$ and wavelength spread $\Delta\lambda / \lambda$. Passing through the sample (S) the neutrons are recorded at the detector (D) as a function time. The solid and dashed lines represent neutron trajectories, where the slope corresponds to the velocity of the neutrons. A periodic perturbation at the sample (green) depicts changes of the sample scattering intensity with period $T_\mathrm{S}$. The oscillation at the detector (red) depicts changes of the recorded signal with a period $T_\mathrm{D} = T_\mathrm{S}$. The gray shaded area represents the distribution of neutron velocities associated with a wavelength spread $\Delta\lambda / \lambda$, which generates a the time-smearing of the detector signal characterized by $\Delta t$.}
\label{fig:sans_timedistance_diagram}
\end{figure}

Conventional SANS with a continuous beam permits to resolve slow changes of the scattering pattern as a function of time. The associated time versus distance diagram is depicted in Fig.~\ref{fig:sans_timedistance_diagram}. A velocity selector (C) transmits a quasi-continuous beam of neutrons with a wavelength $\lambda$ and a wavelength spread $\Delta\lambda / \lambda$. Typical neutron trajectories are represented by the solid and dashed lines where the slope corresponds to the velocity of the neutrons, $v_\mathrm{n}$, and the gray shaded areas reflect the distribution of velocities due the wavelength spread $\Delta\lambda / \lambda$. The neutrons interact with a sample (S) which is subject to a periodic modulation with period of modulation $T_\mathrm{S}$, \textit{e.g.}, due to a AC magnetic field. Here, the oscillation depicted in green at the sample position represents the periodic variation of the scattering by the sample as driven by the external modulation. 

Following the scattering process, the neutrons are recorded at a detector (D) placed at a distance $L_\mathrm{SD}$ behind the sample. As a function of time data recorded at the detector are binned with a period $T_\mathrm{D}$. Neutrons that pass through the sample at a given phase of the sample modulation contribute to the detector signal at the same phase of the oscillation. In the idealized case, when all neutrons have the same velocity, $v_\mathrm{n}$, the time dependence of the detector signal follows the time-dependence of the sample modulation precisely, shifted by the time the neutrons need to travel the distance $L_\mathrm{SD}$. Hence, in order to collect accurate data, the sample modulation and the detector data acquisition systems need to be synchronized such that $T_\mathrm{D} = T_\mathrm{S}$. 

It is now instructive to consider typical velocity selectors used at SANS beam lines, which generate a triangular velocity distribution~\cite{Wagner_PhysB180_1992}. This velocity distribution results in a spread of neutron trajectories as depicted by gray shading in Fig.~\ref{fig:sans_timedistance_diagram}, where the limiting velocities may be denoted $v_\mathrm{min}$ and $v_\mathrm{max}$. For the purpose of the discussion marked in gray shading and bounded by dashed lines are neutrons with velocities that pass through the sample at the same point of time of the modulation. These neutrons will arrive at the detector at different points of time. This causes an averaging of the detector signal as a function of time, where the time difference between the signal produced by neutrons with speed $v_\mathrm{n}$ and those with velocities $v_\mathrm{min}$ or $v_\mathrm{max}$ is given by
\begin{equation}
\Delta t = \frac{L_\mathrm{SD}}{v_\mathrm{n}} \left( \frac{\Delta \lambda}{\lambda} \right)
\label{eqn:sans_dt}
\end{equation}
Inserting values of typical SANS beam lines,  $L_\mathrm{SD} = 10~\mathrm{m}$, $\lambda = 4.5$\,\AA, and $\Delta\lambda / \lambda = 10\,\% $, one finds $\Delta t \approx 1.14\,\mathrm{ms}$. Hence, in order to prevent significant averaging of the data recorded, the sample modulation should be chosen such that $T_\mathrm{S} > 10~\Delta t$, corresponding to frequencies $f_\mathrm{S} < 87.7~\mathrm{Hz}$. Reducing the wavelength spread, $\Delta t$ may be decreased, however, at the expense of neutron flux.

\subsection{Time-resolved SANS experiments with a pulsed beam}

To improve the resolution in time-resolved SANS, with minimal loss in intensity, a pulsed-beam technique proposed by R. G\"ahler~\cite{Kippling_PLA372_2008} known as TISANE may be used. The technique uses interlocking of the phases for neutron pulse, sample modulation, and detector signal binning which allows a broad band of wavelengths to be used without degrading the time resolution.

\begin{figure}
\includegraphics[width=1.0\linewidth]{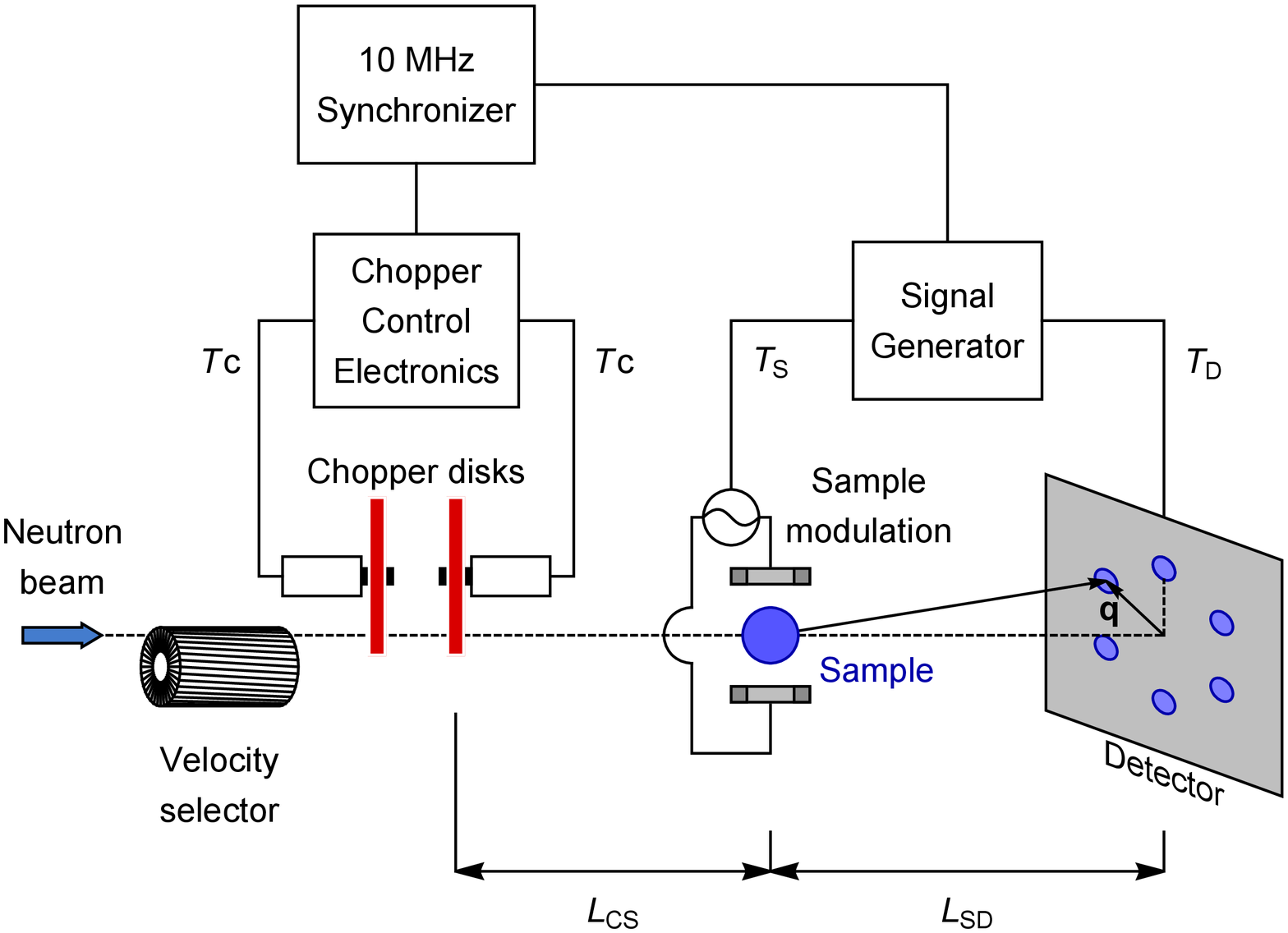}
\caption{Schematic depiction of the experimental setup used for the Time-Involved Small Angle Neutron scattering Experiment (TISANE) as implemented at SANS-1 (FRM II). Following the velocity selector a chopper generates neutron pulses with repetition time $T_\mathrm{C}$. The neutrons pass through a sample subject to a periodic perturbation with period of oscillation $T_\mathrm{S}$. Neutrons are recorded at a detector with with period of time binning $T_\mathrm{D}$. The scattered neutrons are characterised by the scattering vector $\mathbf{q}$. The chopper control electronics, the sample modulation system, and the detector are synchronised by the master trigger generator operating at 10\,MHz.}
\label{fig:tisane_instrument}
\end{figure}

Shown in Fig.~\ref{fig:tisane_instrument} is the schematic layout of a TISANE instrument, \textit{e.g.}, as implemented the SANS-1 beamline at FRM II. A velocity selector produces a continuous neutron beam with a wavelength $\lambda$ and a wavelength spread $\Delta\lambda / \lambda$. Next, a chopper system generates neutron pulses with a repetition time $T_\mathrm{C}$. The system used at SANS-1 consists of two disks placed at 50\,mm distance from each other. Each chopper disk has Boron-covered blades with 14~windows, each window corresponding to an opening angle of $9.06^{\circ}$, and a set of magnetic bearings allowing chopper disk rotational speeds up to 20000~RPM. The system allows to adjust disk rotation speed, direction, and phase between the disks. Careful selection of these three parameters provides access to a wide range of $T_\mathrm{C}$ and allows to tune the neutron pulse width while maintaining the desired pulse repetition time.

Following the chopper, the neutrons pass through the sample which is placed at a distance $L_\mathrm{CS}$ behind the second chopper disk. The sample is subject to a periodic modulation with a period of oscillation $T_\mathrm{S}$, \textit{e.g.}, generated by an AC magnetic field. Typical sample thicknesses are limited to below a few millimeters in order to avoid multiple scattering events. The neutrons are finally recorded with a period of time binning $T_\mathrm{D}$ at a detector placed at a distance $L_\mathrm{SD}$ behind the sample. The resulting data hence form a continuous time-resolved stream that is binned into a discrete number of time frames.

Choosing the parameters $L_\mathrm{CS}$, $L_\mathrm{SD}$, $T_\mathrm{C}$, $T_\mathrm{S}$, and $T_\mathrm{D}$ appropriately, all neutrons from different chopper pulses that arrive at the sample at the same point of time of the sample state oscillation, i.e., at the same phase, will reach the detector within the same phase (corresponding to the detector frequency) irrespective of their wavelengths. This is known as the TISANE condition. It requires that the chopper control electronics, the sample modulation system, and the detector are synchronized at high accuracy, since a small deviation in the period of the oscillation will cause the phase shift to build up with increasing measurement time, averaging the scattering intensities as result. To avoid any problems due to insufficient synchronization, a master trigger unit is used at SANS-1 (also known as Drive Reference Unit or DRU). Equipped with a high precision quartz oscillator, the DRU transmits the reference signal with a base frequency of 10\,MHz to the chopper system acting as a virtual master chopper of the AC magnetic field generator and the detector. As an alternative approach bespoke electronics, referred to as the detector trigger generator (DTG), had been used~\cite{Glinka_2020_JAC_53}. It allows continuous monitoring of $f_\mathrm{S}$ and $f_\mathrm{C}$, and recalculates $f_\mathrm{D}$ required to satisfy the TISANE condition during the measurement. It is also worth mentioning work, where the instrument was used in the Time-of-Flight mode at the spallation neutron source~\cite{Adlmann_2015_JAC_48}. There, data were acquired in the event-mode and synchronized on an absolute time scale with the sample modulation system. After the experiment, the data was processed using the correlation between the neutron events and the oscillatory cycle, and the detector signal rate was adjusted.

\begin{figure}
\includegraphics[width=1.0\linewidth]{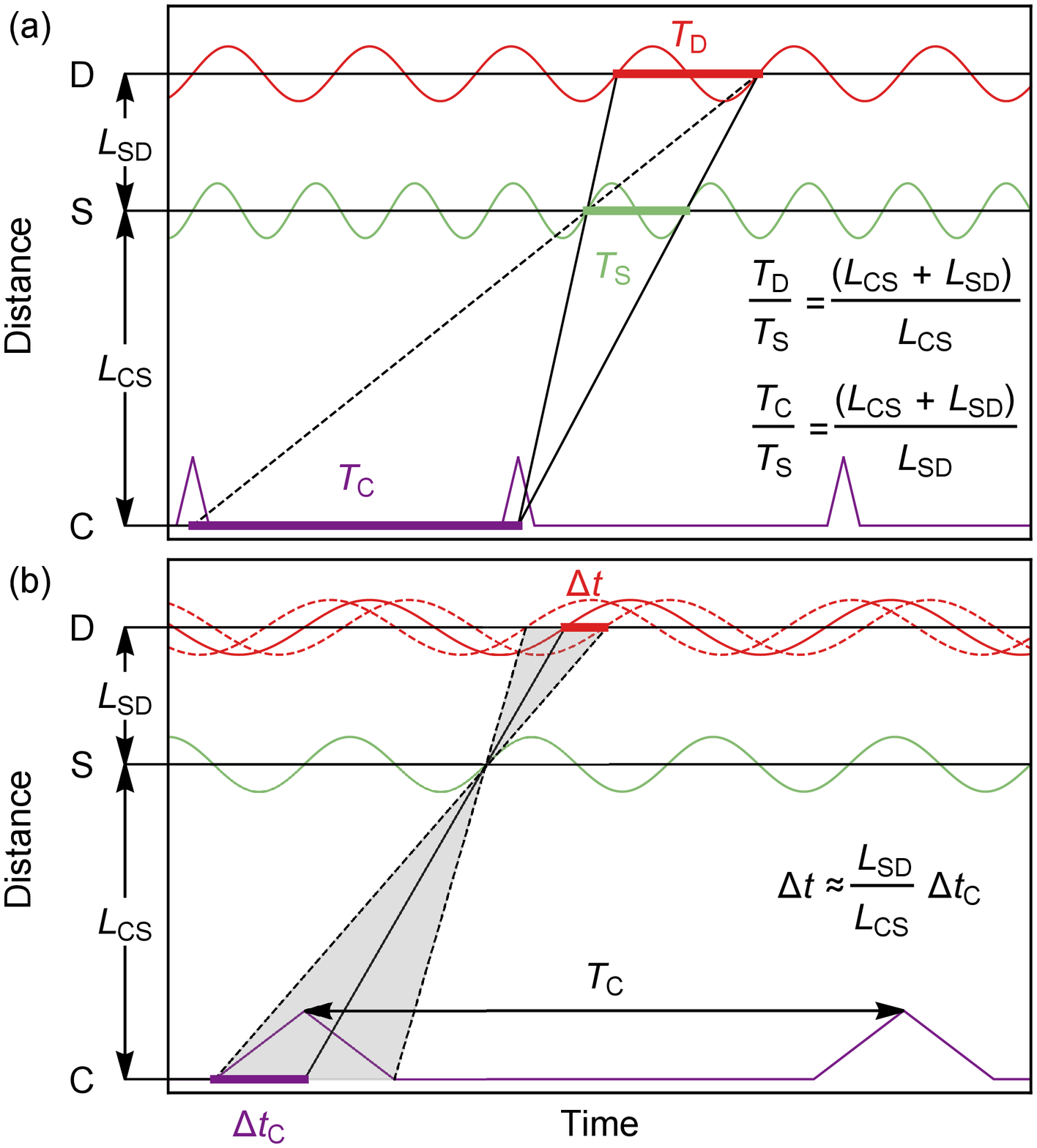}
\caption{Time versus distance diagrams of TISANE. The chopper, sample, and detector are denoted (C), (S), and (D), respectively. Lines represent neutron trajectories, where the slope corresponds to the velocity. A periodic perturbation of the sample (green) will cause a periodic oscillation of the neutron intensity at the detector (red) 
(a) Depiction of the TISANE condition. Neutrons with different velocities that start from the center of a chopper pulse reach the sample and the detector at the same phase of time dependence. (b) Signal smearing due to a finite chopper pulse width. The gray shaded area represents a distribution of neutrons with different velocities which arrive at the sample at the same phase of the modulation. Neutrons starting with the time difference $\Delta t_\mathrm{C}$ with respect to the center of the pulse reach the detector with delay $\Delta t$ causing a smearing.}
\label{fig:tisane_timedistance_diagram}
\end{figure}

The TISANE condition may be illustrated in a time versus distance diagram as shown in Fig.~\ref{fig:tisane_timedistance_diagram}\,(a). The continuous neutron beam is transformed into neutron pulses with repetition time $T_\mathrm{C}$ at the chopper (C). Neutrons passing through the sample (S), which is modulated at a period $T_\mathrm{S}$, are recorded at the detector (D) with a period of time binning $T_\mathrm{D}$. Neutrons that start from the center of a chopper pulse which arrive at the sample at a given phase of the modulation will contribute to the same phase of the detector signal irrespective of the neutron velocity. Graphically this is depicted in Fig.~\ref{fig:tisane_timedistance_diagram}\,(a) for two neutrons starting from the same chopper pulse with different velocities (solid lines). In addition, neutrons from an earlier pulse (dashed line) must reach the sample and the detector also at the correct phase. Corresponding to this condition, the relations between $T_\mathrm{D}$, $T_\mathrm{C}$, and $T_\mathrm{S}$ as determined by the distances between the chopper system, the sample, and the detector are given by
\begin{align}
\frac{T_\mathrm{D}}{T_\mathrm{S}} &= \frac{L_\mathrm{CS} + L_\mathrm{SD}}{L_\mathrm{CS}} \\
\frac{T_\mathrm{C}}{T_\mathrm{S}} &= \frac{L_\mathrm{CS} + L_\mathrm{SD}}{L_\mathrm{SD}}
\label{eqn:tisane_condition}
\end{align}

As presented so far, the TISANE condition assumes infinitely narrow chopper pulses. However, the finite chopper pulse width $\Delta t_\mathrm{C}$ encountered in real experiments results in a distortion of the detector signal as illustrated in Fig.~\ref{fig:tisane_timedistance_diagram}\,(b). Neutrons starting with a time difference from the pulse center will reach the detector shifted by this difference as weighted by $L_\mathrm{SD} / L_\mathrm{CS}$. Additional contributions $\Delta t_\mathrm{S}$ may arise from the averaging of the signal due to the finite flight time of the neutrons across the sample. In a similar manner, a contribution $\Delta t_\mathrm{D}$ may be expected from the finite thickness of the detector. The total time resolution at the detector is then given by~\cite{Wiedenmann_PRL97_2006}
\begin{equation}
\Delta t^{2} = \left( \Delta t_\mathrm{C} \cdot \frac{L_\mathrm{SD}}{L_\mathrm{CS}} \right)^2 + \left( \Delta t_\mathrm{S} \cdot \frac{L_\mathrm{CS} + L_\mathrm{SD}}{L_\mathrm{CS}} \right)^2 + \Delta t^{2}_\mathrm{D}.
\label{eqn:tisane_dt}
\end{equation}

For typical chopper systems the values of $\Delta t_\mathrm{C}$ may vary between $50~\mu$s and $500~\mu$s. In comparison, typical sample thicknesses of a few mm correspond to $\Delta t_\mathrm{S} \approx 5~\mu$s. Typical diameters of the detector tubes of several mm correspond to $\Delta t_\mathrm{D} \approx 10~\mu$s.  As a result, the time resolution of TISANE is essentially determined by the pulse width $\Delta t_\mathrm{C}$.

\subsection{Advantages and limitations or TISANE}

A key advantage of the TISANE technique concerns the possibility to adjust the time resolution and optimize the beam intensity for increasing frequencies of the sample modulation. The details require careful consideration of the chopper duty cycle and frame overlap as discussed in this section. Assuming a chopper system that allows to tune the pulse width independently from the pulse repetition time, it is convenient to define the ratio between the chopper pulse width and repetition time as the chopper duty cycle $D_\mathrm{C}$
\begin{equation}
D_\mathrm{C} = \frac{\Delta t_\mathrm{C}}{T_\mathrm{C}}
\label{eqn:duty_cycle}
\end{equation}
Ignoring minor contributions by $\Delta t_\mathrm{S}$ and $\Delta t_\mathrm{D}$, the time resolution of the instrument may hence be expressed as a function of the duty cycle
\begin{equation}
\Delta t = D_\mathrm{C} T_\mathrm{D} = D_\mathrm{C} \frac{L_\mathrm{CS} + L_\mathrm{SD}}{L_\mathrm{CS}} T_\mathrm{S}.
\label{eqn:tisane_dt_final}
\end{equation}

It is important to keep in mind that the changes of the scattering intensity observed at the detector are characterized by the period of the detector signal oscillation $T_\mathrm{D}$ rather than the period of the sample modulation $T_\mathrm{S}$. Thus, for a given $T_\mathrm{D}$, the signal quality may be inferred from $D_\mathrm{C}$. This way, the frequency range accessible in the sample modulation is essentially limited by the maximum rotational speed of the chopper disks (until $\Delta t_\mathrm{S}$ and $\Delta t_\mathrm{D}$ become comparable to $\Delta t_\mathrm{C}$). The approach is, however, limited by the neutron flux being proportional to the chopper duty cycle $D_\mathrm{C}$. While increasing $D_\mathrm{C}$ past a certain value results in a decrease of the signal contrast due to insufficient time resolution, decreasing $D_\mathrm{C}$ would be on the expense of the resolution and require significantly longer measurement times to compensate for the reduction of intensity. For example, the simulations for the case of the harmonically modulated sample scattering function yield the optimal chopper duty cycle value of 11\,\%~\cite{Kippling_PLA372_2008}.

In addition, it is also possible to maximise the intensity by exploiting the spread of neutron velocities, since the time resolution remains unaffected as long as the TISANE condition is satisfied. Namely, when the spread of neutron velocity is sufficiently large, overlap between consecutive neutron pulses is reached such that neutrons that originate in different pulses pass through the sample at the same point of time. The number of chopper openings contributing to the sample intensity at a given instant in time is referred to as overlap factor $N_\mathrm{OF}$, given by
\begin{equation}
N_\mathrm{OF} \cdot T_C = \frac{L_\mathrm{CS}}{v_\mathrm{min}} - \frac{L_\mathrm{CS}}{v_\mathrm{max}}.
\label{eqn:frame_overlap}
\end{equation}
For $N_\mathrm{OF} < 1$ the flux of neutrons at the sample is not constant. In this limit the detector signal may exhibits parasitic contributions due to, e.g., residual Fourier terms of the chopper transmission function and the fundamental chopper and sample frequencies. 

The scattering intensity recorded at the detector may be calculated as the product of the neutron velocity distribution function, $F$, the chopper transmission function, $P_\mathrm{C}$, and the sample scattering function, $S$, integrated over all possible neutron velocities~\cite{Kippling_PLA372_2008}
\begin{equation}
I_\mathrm{D}(t, q) = I_0 \int^{\infty}_{0} F(v) \cdot P_\mathrm{C}(t_\mathrm{C}) \cdot S(t_\mathrm{S}, q)~dv,
\label{eqn:detector_signal}
\end{equation}
where the time at the chopper and the sample is corrected by the offsets $t_\mathrm{C} = t - (L_\mathrm{CS} + L_\mathrm{SD})/v$ and $t_\mathrm{S} = t - (L_\mathrm{SD})/v$, respectively. The detector signal is here calculated neglecting variations of flight time due to the beam divergence and due to small angle scattering. 

\begin{figure}
\includegraphics[width=1.0\linewidth]{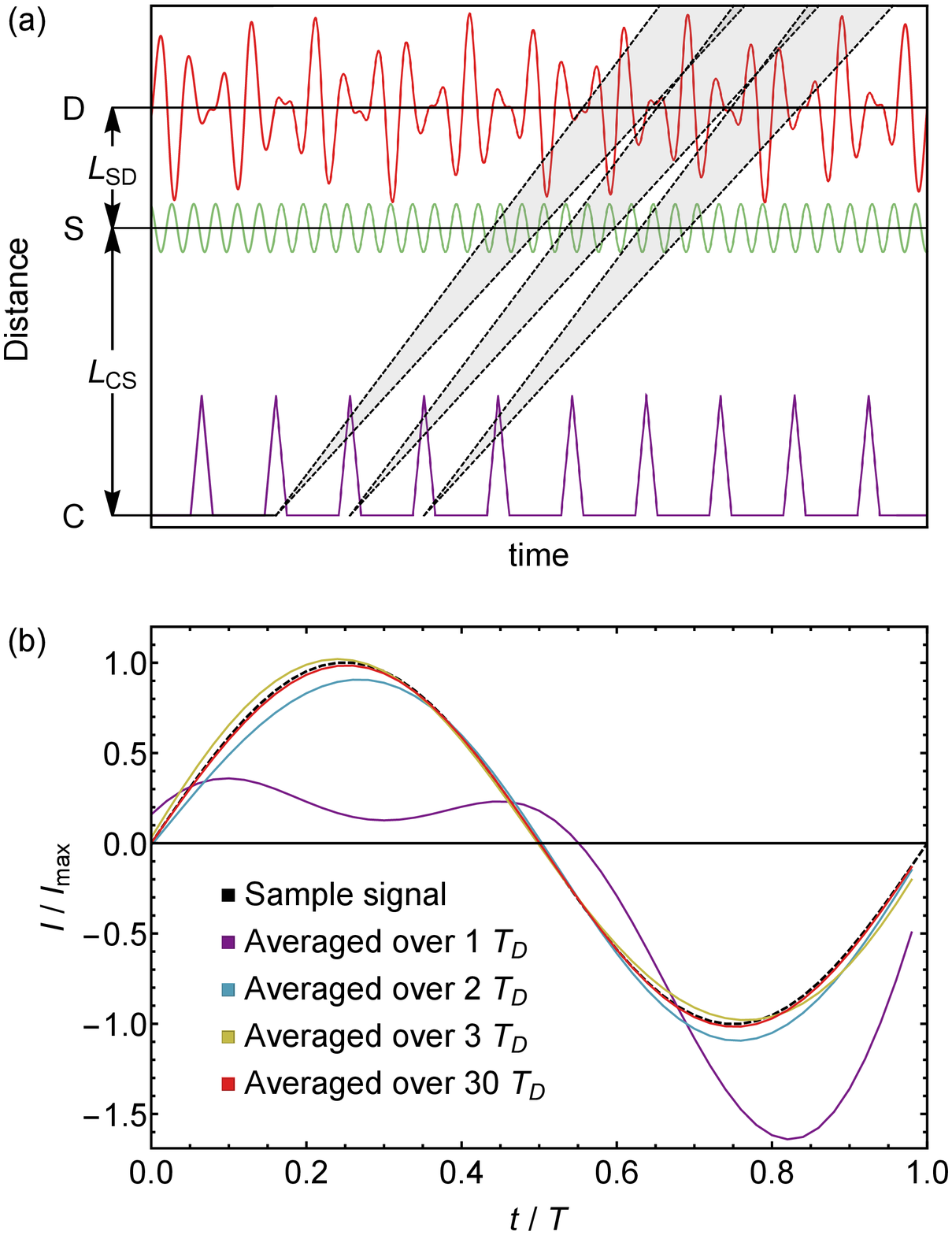}
\caption{Effect of the low frame overlap on the detector signal. (a) Time versus distance diagram for $N_\mathrm{OF} \approx 0.65$. Insufficient frame overlap results in a strongly varying beam intensity at the sample, causing additional components in the detector signal. (b) Averaged detector signal for different measurement durations. The dashed black line represents the original sample signal. The coloured lines show detector signals averaged over multiple periods $T_\mathrm{D}$ as stated in the legend.}
\label{fig:frame_overlap}
\end{figure}

An example for the intensities expected is depicted in Fig.~\ref{fig:frame_overlap}\,(a). The distribution of neutron velocities was modelled by a triangular shape with $\lambda = 4.5\,$\AA~and $\Delta \lambda / \lambda = 10\,\%$. The triangular chopper pulse function with $D_\mathrm{C} = 0.15$ was approximated by a Fourier series. For modelling the oscillation of the scattering intensity of the sample a harmonic motion was assumed with a frequency $f_\mathrm{S} = 403.7\,\mathrm{Hz}$. The distances $L_\mathrm{CS} = 23.9\,\mathrm{m}$ and $L_\mathrm{SD} = 10.0\,\mathrm{m}$ were chosen in agreement with the configuration of the experiment discussed below and the frame overlap was $N_\mathrm{OF} \approx 0.65$. 

Keeping the overlap factor $N_\mathrm{OF}$ well above unity should reduce the presence of undesired frequencies in the detector signal, in a typical TISANE configurations it may be achieved at the values exceeding $N_\mathrm{OF} > 10$~\cite{Kippling_PLA372_2008}. If sufficiently large values of $N_\mathrm{OF}$ are not accessible (\textit{e.g.} for the low sample modulation and hence chopper frequencies), parasitic signal components may be filtered out during the process of the detector intensity averaging, calculated as a sum over several time constants $T_\mathrm{D}$

\begin{equation}
I_\mathrm{avg}(t) = \frac{1}{n_\mathrm{max}} \sum^{n_\mathrm{max}}_{n = 0} I_\mathrm{D} (t + n T_\mathrm{D}).
\label{eqn:signal_avg}
\end{equation}

The results of such an averaging are illustrated in Fig.~\ref{fig:frame_overlap}\,(b). It is worth mentioning that in a typical TISANE, data are always collected over a large number of detector periods for the sake of intensity, which would naturally filter out these ``parasitic'' signal components. For the reference considered here, the measurement times required for the experimental data shown in Fig.~\ref{fig:time_to_rock} were 120\,s, with the $f_\mathrm{D} = 284.7075\,\mathrm{Hz}$ giving $n_\mathrm{max} \approx 3.4 \cdot 10^4$.

To summarize, the typical time resolution of TISANE is between $0.02$\,ms and $2.00$\,ms. For example, SANS-1 permits studies at sample oscillation frequencies up to $f_\mathrm{S} \approx 30\,\mathrm{kHz}$. The technique is designed to probe periodic changes in the neutron scattering pattern, but does not permit to resolve individual stochastic events. While there are no strict limitations at lower frequencies, the TISANE advantage of time resolution being proportional to the period of detector signal oscillation becomes less prominent. Considering the intensity losses at the chopper system, one may argue that conventional time-resolved SANS with a continuous beam might be preferable for sample frequencies up to few hundreds of Hz. A precise comparison of the techniques may be possible when comparing Eq.~\ref{eqn:sans_dt} to Eq.~\ref{eqn:tisane_dt_final}. It would depend on various details of implementation, notably beam intensity, desired $q$-resolution, range of available neutron wavelengths \textit{etc}. 

\section{Parasitic signal contributions}
\subsection{TISANE of the skyrmion lattice (SL) motion in chiral magnets}

In the following we present TISANE data recorded in a kinetic neutron scattering study of the skyrmion lattice (SL) motion in chiral magnets. Skyrmions are topologically non-trivial spin textures that exhibit an exceptionally efficient coupling to spin currents, notably spin-polarized charge currents and magnon currents~\cite{Schulz_NatPhys8_2021, Everschor_PRB86_2012, Mochizuki_NatMat13_2014, Zhang_NatComm9_2018}. The data we report here follow up on an investigation carried out at the SANS beamline V4 at HZB \cite{mue_NJP2016}. The work reported here concerned the information content and putative presence of parasitic signal contributions at large excitation amplitudes, when the skyrmion lattice follows the oscillatory motion of the field direction.

\begin{figure}
\includegraphics[width=1.0\linewidth]{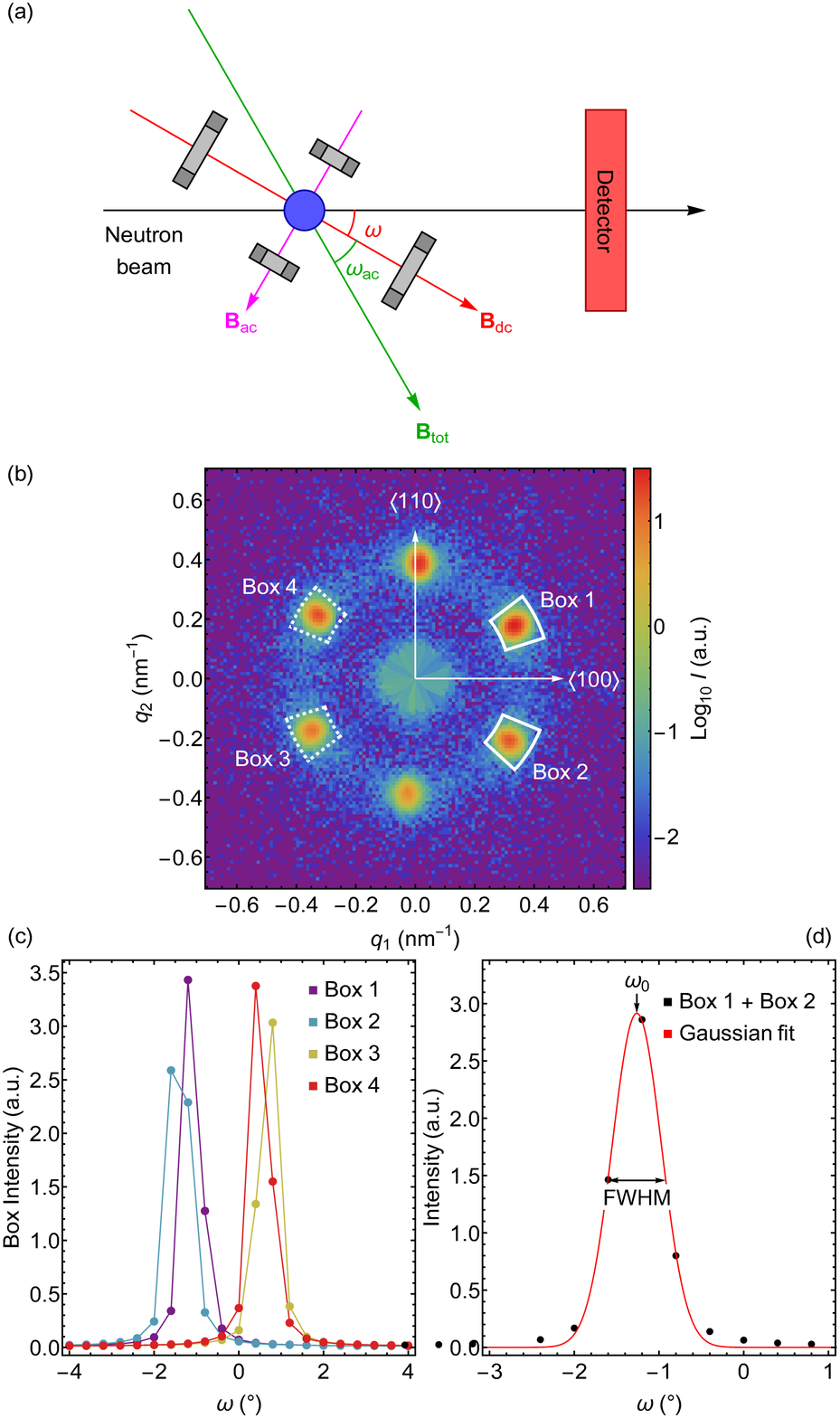}
\caption{Basic aspects of TISANE in the skyrmion lattice of MnSi. (a) The static magnetic field $\mathbf{B}_\mathrm{dc}$ and the oscillating magnetic field $\mathbf{B}_\mathrm{ac}$ were generated by Helmholtz coils depicted in gray shading. The orientation of the resulting magnetic field $\mathbf{B}_\mathrm{tot}$ with respect to $\mathbf{B}_\mathrm{dc}$ is denoted by the angle $\omega_\mathrm{ac}$. The orientation of the static field with respect to the incident neutron beam is denoted by the rocking angle $\omega$.
(b) Typical SANS intensity pattern of the skyrmion lattice (SL) in MnSi as stabilized by a static magnetic field $\mathbf{B}_\mathrm{dc}$ (no $\mathbf{B}_\mathrm{ac}$ present). Crystallographic  $\langle 110 \rangle$ and $\langle 100 \rangle$ axes were vertical and horizontal, respectively. Boxes 1, 2, 3, 4 denote detector segments in which scattering intensity was observed.
(c) Typical integrated scattering intensity recorded in the boxes marked in panel (b) as a function of $\omega$. 
(d) Average intensity recorded in Boxes 1 and 2. The red line represents a Gaussian fit of the data, with $\omega_0$ denoting the rest position of the skyrmion lattice in the absence of an AC field.}
\label{fig:sans_skx}
\end{figure}

The experiment was carried out at the beam-line SANS-1 at FRM-II, Garching, Germany~\cite{Muehlbauer_NIMPRA297_2016}. Unpolarized neutrons were used with a wavelength of $\lambda = 4.5 \,$\AA~and a wavelength spread $\Delta\lambda / \lambda \sim 10 \, \%$ (FWHM). The collimation distance was set to 12\,m, the sample-detector distance was $L_\mathrm{SD} = 10.025\,\mathrm{m}$, and the chopper-sample distance was $L_\mathrm{CS} = 23.925\,\mathrm{m}$.  The chopper system was operated in a mode where one chopper disk was spinning while the second chopper disk was kept in a fixed position. With each chopper disk having 14 openings and each opening having an angular width of $9.06^{\circ}$, this resulted in a chopper duty cycle of $D_\mathrm{C} = 0.352$.

A spherical single crystal of MnSi with a diameter of 5.8\,mm was studied. The spherical sample shape served to ensure uniformity of both demagnetizing and hence internal fields~\cite{Adams_PRL107_2011}. The oscillatory motion of the direction of the applied magnetic field was generated by means of the superposition of crossed DC and AC fields, each produced by a set of Helmholtz coils. The static magnetic field was $B_\mathrm{dc} = 170\,\mathrm{mT}$. The amplitude of the oscillating magnetic field was $B_\mathrm{ac} = 6.11\,\mathrm{mT}$ at a frequency $f_\mathrm{S} = 403.7075\,\mathrm{Hz}$. In order to account for shielding of the AC magnetic field by the cryostat and instrument components, the magnetic field was calibrated with a Hall probe at room temperature. The sample was cooled to a temperature $T = 28.0\,\mathrm{K}$ by means of a closed cycle cryostat equipped with a quartz vacuum shield. Additional heating effects due to the AC field were carefully compensated. 

A schematic depiction of the experimental setup is shown in Fig.~\ref{fig:sans_skx}\,(a), featuring the direction of the incident neutron beam, the sample, the magnetic static and oscillating magnetic fields $\mathbf{B}_\mathrm{dc}$ and $\mathbf{B}_\mathrm{ac}$, and the detector. A static field, $\mathbf{B}_\mathrm{dc}$, needed to stabilize the skyrmion lattice phase was generated by a set of Helmholtz coils depicted in gray shading. The direction $\mathbf{B}_\mathrm{dc}$ is tilted with respect to the incident neutron beam by the rocking angle $\omega$. The oscillation of the direction of the magnetic field was generated with a small AC field, $\mathbf{B}_\mathrm{ac}$, aligned perpendicular to the dc-field. The direction of the resulting total magnetic field $\mathbf{B}_\mathrm{tot}$ with respect to $\mathbf{B}_\mathrm{dc}$ is characterized by the angle $\omega_\mathrm{ac}$. The sample was aligned in a way that the vertical axis of the system was parallel to a crystallographic $\langle 110 \rangle$ direction, and the direction of the dc-field $\mathbf{B}_\mathrm{dc}$ coincided with another $\langle 110 \rangle$ crystallographic direction.

Shown in Fig.~\ref{fig:sans_skx}\,(b) is a typical six-fold intensity pattern of the skyrmion lattice in the absence of the AC field~\cite{Muehlbauer_Sci323_2009}. The pattern was recorded at $\omega = -0.4^{\circ}$ corresponding to the zero position. The diffuse intensity contributions in the background are visible on a logarithmic scale only; they are very weak and may be ignored in what follows. For the TISANE four peaks were selected as marked by the boxes denoted 1 through 4. The integrated intensity in these boxes as a function of rocking angle is shown in Fig.~\ref{fig:sans_skx}\,(c). The two remaining peaks at the top and the bottom are not tracked as they do not depend on $\omega$. 

Comparison of the intensities in the four boxes underscores the sensitivity of the scattering intensity to the precise orientation between the sample, neutron beam and the magnetic field. In case of a perfectly aligned system, Figure~\ref{fig:sans_skx}\,(c) should exhibit symmetry around $\omega = 0$: curves for boxes 1 and 4 and boxes 2 and 3 should cross at $\omega = 0$, and respective curves for boxes 1 and 2 and boxes 3 and 4 should overlap. To compensate for the small misalignment, we take the sum of the intensity of box 1 and box 2 as a function of $\omega$ as shown in Fig.~\ref{fig:sans_skx}\,(d). The resulting curve may be described well with a Gaussian featuring a full width at half maximum $\mathrm{FWHM} = 0.68^{\circ}$. The shape and the width of the curve contained the information about the mosaicity of the SL structure. The area under the curve is proportional to the volume fraction of SL phase. The position of the peak center $\omega_0 = -1.27^{\circ}$ represents the orientation of the SL in the absence of AC magnetic field. Tracking deviations of the peak position permitted to track changes of the SL direction. 

\begin{figure}
\includegraphics[width=1.0\linewidth]{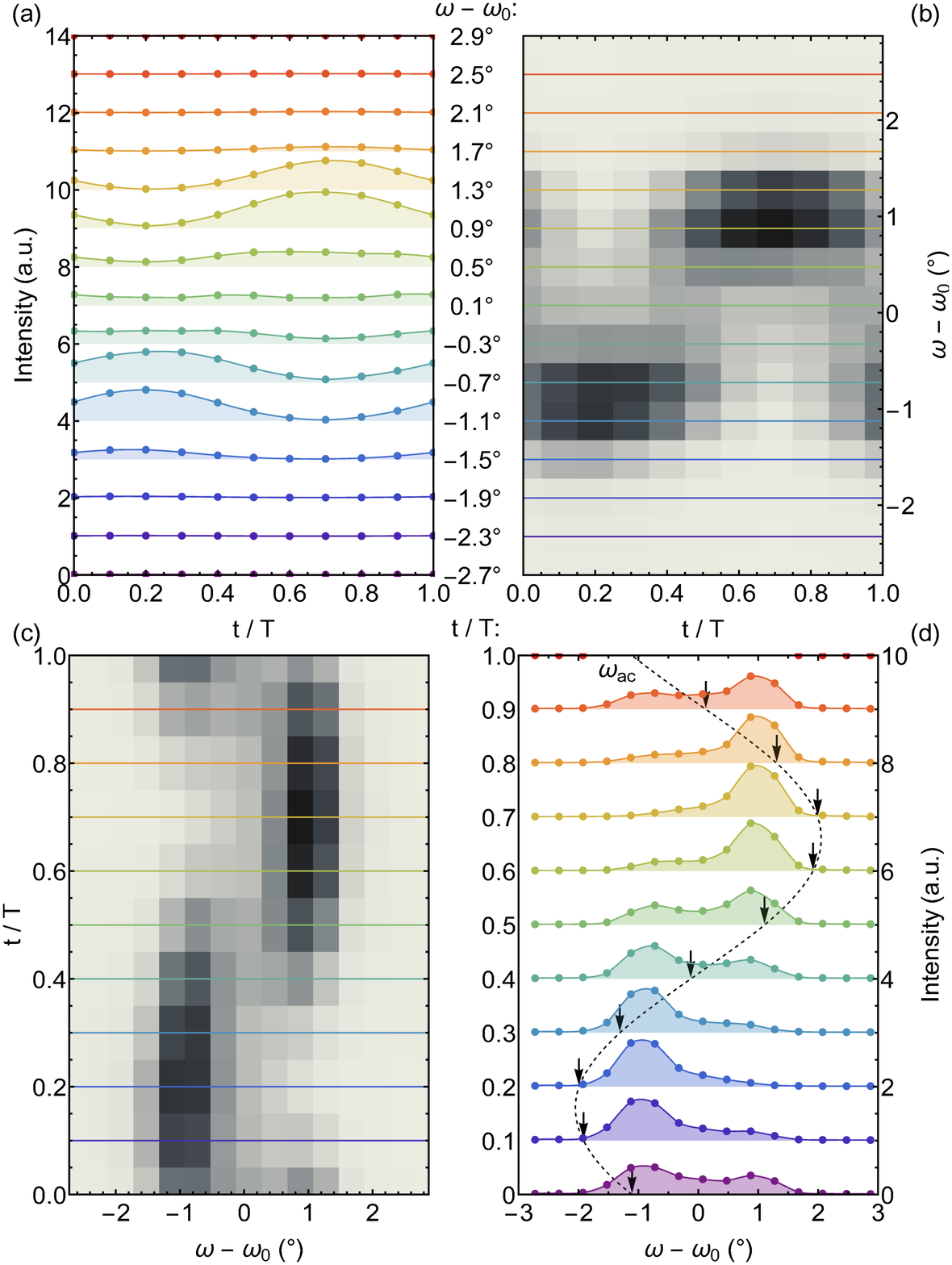}
\caption{Acquisition process of time-resolved rocking scans. (a) Sum of the intensities in box 1 and box 2 in \textit{cf.} Fig.~\ref{fig:sans_skx}) as a function of time $t / T$ for fixed positions of rocking angle $\omega - \omega_0$ as denoted on the right hand side. Data have been shifted vertically for clarity. (b) Two-dimensional time-resolved rocking map inferred from panel (a). Horizontal lines correspond to specific rocking angles where the color coding is that of panel (a). (c) Transposed two-dimensional intensity map of the map shown in panel (b). Horizontal lines denote time frames. (d) Rocking peaks for different time frames. The dashed line represents the variation of the field direction $\omega_\mathrm{ac}$. The arrows mark the value of $\omega_\mathrm{ac}$ for each respective time bin. Data are shifted vertically for clarity.}
\label{fig:time_to_rock}
\end{figure}

To determine time-dependent rocking scans we recorded at first the scattering intensity for selected fixed rocking angles $\omega - \omega_0$ as a function of time. Data were then sorted into a number of intervals according to different points of time during a full period of the oscillation. Note that the data were recorded at the detector with the period of oscillation $T_\mathrm{D}$, by plotting the intensity as a function $t / T_\mathrm{D}$ it is possible to relate the intensity variations to the modulation of the sample state characterized by $t_\mathrm{S} / T_\mathrm{S}$, therefore we omit the notations and write it as $t / T$. The sum of the intensities of box 1 and 2 (see Fig.~\ref{fig:sans_skx}\,(b)) as a function of time is shown in Fig.~\ref{fig:time_to_rock}\,(a). The color shading denotes different rocking angles $\omega - \omega_0$. Based on these data the two-dimensional time-resolved rocking map was generated shown in Fig.~\ref{fig:time_to_rock}\,(b). Here horizontal lines correspond to specific rocking angles where the colours correspond to Fig.~\ref{fig:time_to_rock}\,(a). 

In order to extract the rocking curves for a sequence of different time frames, the two-dimensional rocking map was transposed as shown in Fig.~\ref{fig:time_to_rock}\,(c). This way data are depicted as a function of $t / T$. The horizontal lines mark time frames rather than rocking angles. The resulting rocking curves are depicted in Fig.~\ref{fig:time_to_rock}\,(d) where the coloured lines denote the time frames marked in Fig.~\ref{fig:time_to_rock}\,(c). Taken together these time-resolved TISANE rocking curves are the result of a comprehensive data set, noting that a correct arrangement of time and rocking angles is essential. 

At first sight Fig.~\ref{fig:time_to_rock}\,(d) appears to show that the SL follows the direction of $B_\mathrm{tot}$, where the instantaneous field direction is marked by an arrow. In comparison to a the Gaussian rocking curve observed without oscillation of the field direction shown in Fig.~\ref{fig:sans_skx}\,(d), the intensities are, however, heavily broadened seemingly shifting between two prominent maxima. As discussed in the following, both the qualitative and quantitative form of the TISANE rocking curves are purely parasitic effects that originate in the specific choice of parameters of data binning and pulse length. 

\subsection{Consequences of data binning and pulse length}
\label{subsection:time_resolution}

\begin{figure}
\includegraphics[width=1.0\linewidth]{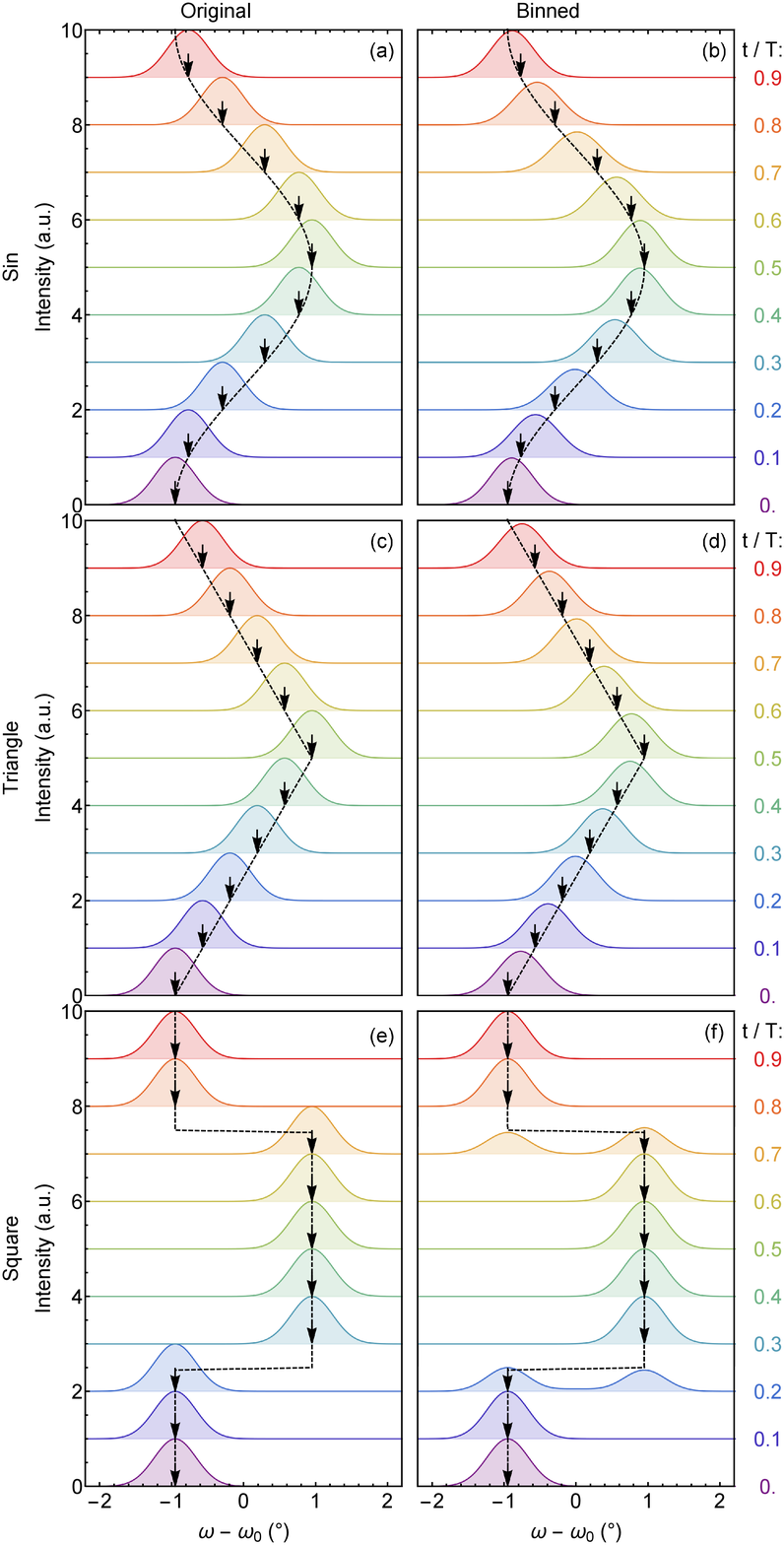}
\caption{Changes of simulated TISANE rocking intensities due to binning in time frames. Data have been shifted vertically for clarity. Panels on the left hand side assume a Gaussian intensity distribution that follows accurately a given periodic time dependence without binning. Panels on the right hand side display the effects of time smearing when binning data in ten time frames. Color shading depicts specific time frames. (a, b) Behaviour for a sinusoidal oscillation. The binning results in a broadening and concomitant reduction of peak height.  (c, d) Behaviour for a triangular motion. The binning essentially results in the same broadening and peak reduction for all time frames. (e, f) Behaviour for a rectangular oscillation. A double peak emerges for time frames of fast changes.}
\label{fig:model_motion}
\end{figure}

To gain insight on the origin of the TISANE rocking curves shown in Fig.~\ref{fig:time_to_rock}\,(d) different basic responses of the SL motion were calculated assuming that the whole SL volume simultaneously and instantly followed changes of the magnetic field direction as described by $\omega_\mathrm{ac}$  (Fig.~\ref{fig:model_motion}). In addition it was assumed that the SL order is characterized by a Gaussian distribution that is not affected by the motion. Quantitative values of the Gaussian rocking intensity as well as the size of $\mathbf{B}_\mathrm{ac}$ corresponded to those observed in the study of MnSi reported in Fig.~\ref{fig:sans_skx} and Fig.~\ref{fig:time_to_rock}. 

It proved to be instructive to consider at first three fundamental oscillatory time-dependences, namely a sinusoidal, a triangular, and a square motion, depicted in the top, middle and bottom row of Fig.~\ref{fig:model_motion}, respectively. As shown in the panels on the left hand side of Fig.~\ref{fig:model_motion}, without binning the data as a function of time, the Gaussian distributions follow any changes of field direction instantly without change of width or intensity. In comparison, panels on the right hand side of  Fig.~\ref{fig:model_motion} depict TISANE rocking scans, when binning the time dependence into ten time frames. Within each time frame, $t_{i}$, all scattering data were thereby integrated from $t_{i}$ to $t_{i+1}$: 
\begin{equation}
I(\omega - \omega_0) |_{t = t_i} = \frac{1}{t_{i+1} - t_{i}} \int^{t_{i+1}}_{t_{i}} I(\omega - \omega_0, \tau)~d\tau
\label{eqn:time_binning}
\end{equation}

As evident in the panels on the right hand side of Fig.~\ref{fig:model_motion}, the binning generates an averaging and concomitant smearing. For the sinusoidal motion, shown in Fig.~\ref{fig:model_motion}\,(b), both the peak height and peak width vary periodically as a function time, where the height and width scale with the rate of change of the field orientation. Accordingly, for the triangular time-dependence shown in Fig.~\ref{fig:model_motion}\,(d) the smearing is essentially the same for all time frames, as the absolute value of the rate of change of the field direction is essentially constant. Contributions by points of time where changes of the direction of motion take place are averaged out. The rectangular motion, finally, exhibits a broadening and a double peak structure in the vicinity of the points of time of instant changes of field orientation only as shown in Fig.~\ref{fig:model_motion}\,(f). While the binning into time-frames does not affect the appearance of the TISANE rocking scans radically, the effect is inevitable and important to keep in mind. Indeed, the impact of the binning may be reduced by increasing the number of time frames, however, at a significant cost of intensity for each time frame. It seems logical to select the number of time bins $N_\mathrm{bin}$ at least high enough so that the duration a single time bin $T_\mathrm{D} / N_\mathrm{bin}$ is less than the instrumental resolution defined in Eq.~\ref{eqn:tisane_dt_final}.

Comparing the calculated rocking scans shown in Fig.~\ref{fig:model_motion}\,(b) with the experimental data shown in Fig.~\ref{fig:time_to_rock}\,(d), the general trend of a broadening that scales with the rate of change of the field direction under sinusoidal motion cannot be ruled out. Yet, the experimental data exhibit a pronounced double peak structure with maxima that are essentially located at two fixed rocking angles $\omega -\omega_0\approx \pm 1^{\circ}$ and scattering intensity that appears to be shifted between these two maxima as a function of time.

\begin{figure}
\includegraphics[width=1.0\linewidth]{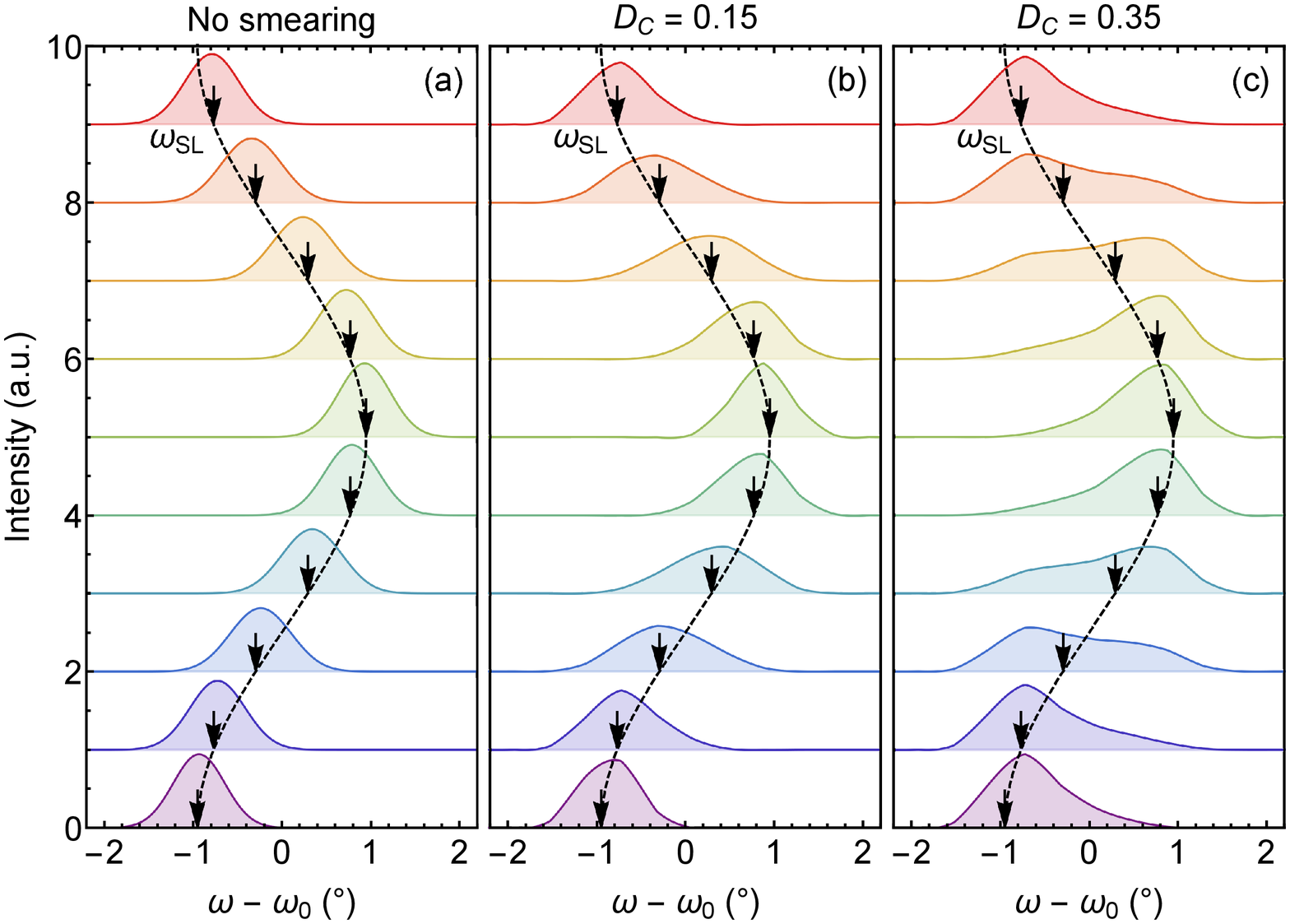}
\caption{Changes of simulated TISANE rocking intensities due to binning for different duty cycles. Data have been shifted vertically for clarity. (a) Gaussian rocking curves that follow a sinusoidal driving field without binning and time smearing. (b) Gaussian rocking curves that follow a sinusoidal oscillation taking into account a TISANE duty cycle $D_\mathrm{C} = 0.15$ and binning into ten time frames. A broadening and a reduction of the Gaussian distribution appears to be present. (c) Gaussian rocking curves that follow a sinusoidal oscillation taking into account a TISANE duty cycle $D_\mathrm{C} = 0.35$ and binning into ten time frames. Besides a seeming broadening and reduction of the Gaussian distribution a double peak structure emerges due to the binning.}
\label{fig:timesmearing_Dc}
\end{figure}

It transpires that the double peak structure may be fully attributed to the choice of neutron pulse length affecting the time resolution defined in Eq.~\ref{eqn:tisane_dt_final}. Considering a chopper made of two disks with rectangular windows a triangular pulse shape with duty cycles typically ranging between 0.10 and 0.35 is obtained. In case of a chopper pulse that is too wide, details of the signal will get heavily averaged out and the shape of the TISANE rocking scans will become symmetric. For a detailed assessment of the influence of the pulse shape and duty cycle on the appearance of the TISANE rocking scans, the intensities may be estimated with Eq.~\ref{eqn:detector_signal}. For the calculations, we assume the TISANE condition is strictly followed, hence the wavelength spread does not impact the time smearing of the detector signal. Additionally we consider the measurement times to be long enough so that $n_\mathrm{max} \to \infty$ when the averaged detector signal is calculated according to Eq.~\ref{eqn:signal_avg}. With that, in the calculations the wavelength spread can be taken as delta function, and the TISANE rocking intensity may then be estimated as convolution of the chopper pulse function, $P_\mathrm{C}$, with the sample scattering function, $S$, namely
\begin{equation}
I_\mathrm{avg}(\phi) = I_0 \int^{2\pi}_{0} P_\mathrm{C}(\Delta \phi) \cdot S(\phi + \Delta \phi)~d\Delta \phi.
\label{eqn:smearing_model}
\end{equation}

Shown in Fig.~\ref{fig:timesmearing_Dc} is an evaluation of the influence of the pulse length for chopper duty cycles $D_\mathrm{C}=0.15$ and 0.35. For ease of comparison shown in Fig.~\ref{fig:timesmearing_Dc}\,(a) is the behaviour without binning and a vanishingly small value of $D_\mathrm{C}$. In this limit the TISANE rocking intensity follows any changes in $\omega_\mathrm{ac}$ while maintaining the peak shape and height as already shown in Fig.~\ref{fig:model_motion}\,(a). Considering now $D_\mathrm{C}=0.15$ the combined effect of binning and finite pulse length causes an averaging akin to Fig.~\ref{fig:model_motion}\,(b). Finally, for $D_\mathrm{C} = 0.35$, the calculated TISANE rocking curves shown in Fig.~\ref{fig:timesmearing_Dc}\,(c) exhibit the double peak behaviour observed experimentally. As the experimental data shown in Fig.~\ref{fig:time_to_rock}\,(d) was measured for a chopper duty cycle $D_\mathrm{C} = 0.352$ this provides a full account of the double peak structure.

\begin{figure}
\includegraphics[width=1.0\linewidth]{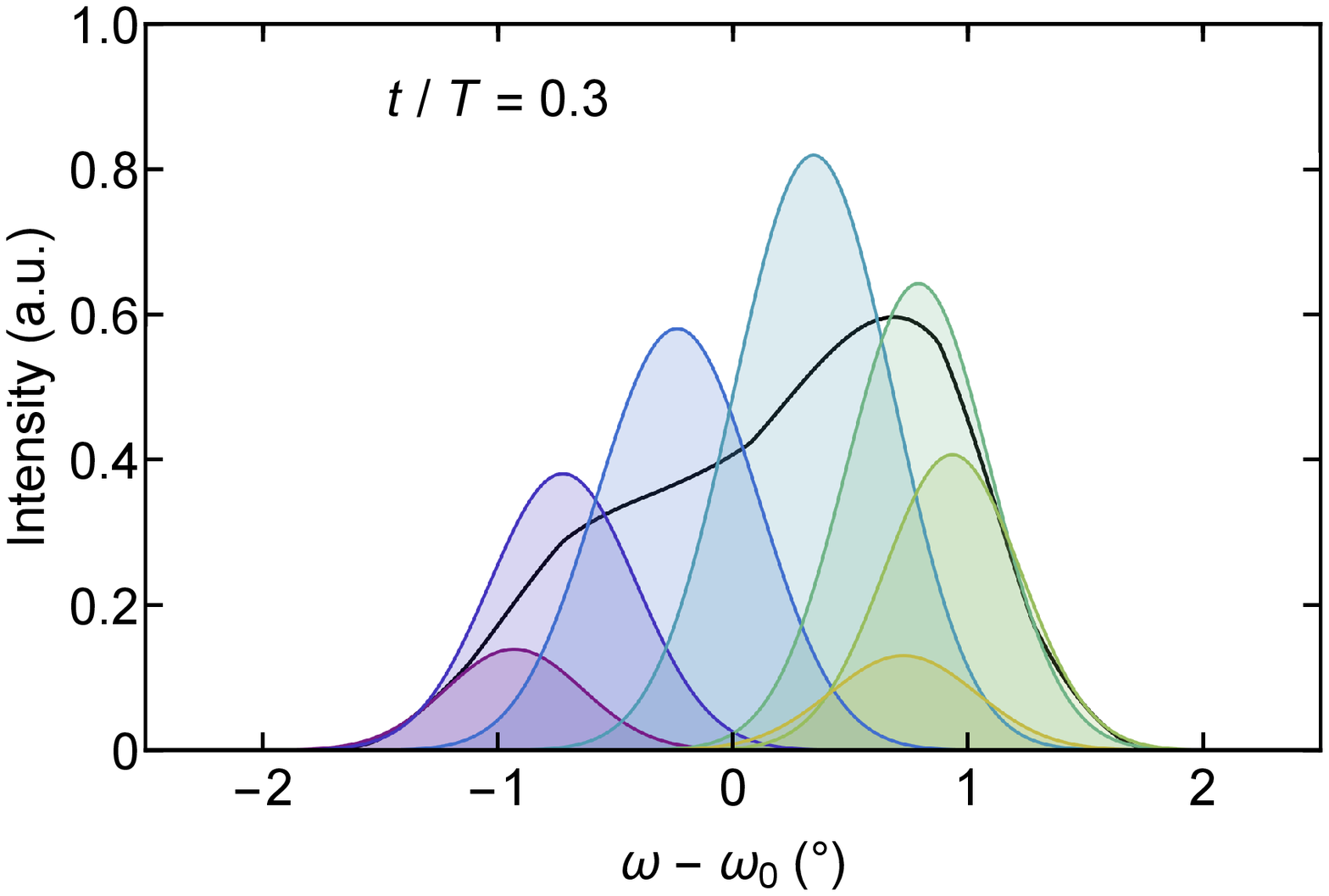}
\caption{Origin of the double peak structure in TISANE rocking scans for a long neutron pulse corresponding to a duty cycle of $D_\mathrm{C} = 0.35$. The black line represents the TISANE rocking scan shown in Fig.~\ref{fig:timesmearing_Dc}\,(c) for a time frame $t / T = 0.3$. Color-shaded Gaussians represent scattering contributions scaled by the chopper pulse function as associated with different time frames due to the long chopper pulse.
}
\label{fig:timesmearing_convolution}
\end{figure}

On an intuitive level the double-peak distribution may be understood as a parasitic superposition of rocking curves due to the wide chopper pulse width. Shown by the black line in Fig.~\ref{fig:timesmearing_convolution} is the TISANE rocking curve displayed in Fig.~\ref{fig:timesmearing_Dc}\,(c) for $t / T = 0.3$. This curve may be understood as a normalized sum of Gaussian intensity profiles that originate at different times within the pulse. The color shading here corresponds to the time frames shown in Fig.~\ref{fig:timesmearing_Dc}\,(c). In other words, the pulse width effectively includes contributions associated with several time frames weighted by the triangular pulse shape at the respective moments of time. 

\section{Summary}
\\

In summary, we reviewed optimization strategies of TISANE in kinetic SANS studies of mesoscale textures. To illustrate parasitic effects we considered the motion of the SL in MnSi under oscillations of the field direction, focusing on the emergence of pronounced qualitative and quantitative changes of the intensity distribution, notably strong velocity dependent broadening and a double peak structure suggestive of periodic shifts of scattering intensity. Simulating the effects of binning into time frames for different duty cycles for a Gaussian rocking distribution that is driven by a sinusoidal, triangular, and square wave excitation, we reproduce the behaviour observed experimentally. This illustrates the potential of TISANE to obtain time-resolved information at high scattering intensities, while emphasizing the need for carefully simulating and choosing key parameters of data detection and analysis. 


\appendix


\ack{Acknowledgements}
\\
We wish to thank A. Bezvershenko, P. B\"oni, A. Heinemann, A. Rosch, and U. Keiderling for fruitful discussions and the staff at the Heinz Maier-Leibnitz Zentrum (MLZ) for support. This work has been funded by the Deutsche Forschungsgemeinschaft (DFG, German Research Foundation) under TRR80 (From Electronic Correlations to Functionality, Project No. 107745057, Project E1), the priority program SPP 2137 (Skyrmionics) under grant PF393/19 (project-id 403191981), and the excellence cluster MCQST under Germany's Excellence Strategy EXC-2111 (Project No. 390814868). Financial support by the European Research Council (ERC) through Advanced Grants No. 291079 (TOPFIT) and No. 788031 (ExQuiSid) is gratefully acknowledged.

\referencelist[tisane]


\end{document}